\documentclass[10pt,a4paper]{extarticle}
\usepackage{graphicx}
\usepackage{xcolor}
\usepackage{geometry}
\usepackage{amsmath,amssymb,amsthm}
\usepackage{booktabs}
\usepackage{tabularx}
\usepackage{enumitem}
\usepackage{caption}
\usepackage{microtype}
\usepackage{adjustbox}
\usepackage{ragged2e}
\usepackage{multicol}
\usepackage{placeins}
\usepackage[bottom,hang,flushmargin]{footmisc}
\usepackage[
    colorlinks=true,
    linkcolor=blue!45!black,
    citecolor=blue!40!black,
    urlcolor=blue!50!black,
    bookmarks=true,
    bookmarksnumbered=true,
    unicode=true
]{hyperref}

\usepackage[numbers,sort&compress]{natbib}
\newtheorem{proposition}{Proposition}
\newtheorem{remark}{Remark}

\newcolumntype{Y}{>{\RaggedRight\arraybackslash}X}

\title{\bfseries The Generalised Kessler Syndrome and the Persian Gulf}
\author{Bernhard K Meister\thanks{bernhard.k.meister@gmail.com}}
\date{\today}

\begin{document}
\maketitle
\thispagestyle{empty}

\begin{quote}
\itshape\small
``My grandfather rode a camel, my father rode a camel, I drive a Mercedes, ..., but my great-great-grandson will ride a camel.''
\par\medskip
\upshape\normalsize
Apocryphal, and maybe optimistic, quote attributed to Sheikh Rashid bin Saeed Al Maktoum.
\end{quote}

\begin{abstract}
\noindent
The term Kessler syndrome was coined for a runaway orbital cascade creating a high-density debris field.
The idea is generalised and applied to supply chains.  
An additive channel enables failures to generate further failures; a subtractive channel erodes the capacity to contain and restore.
When damage outpaces repair, failure becomes irreversible and the system settles into a new state. 

\noindent
Supplies connect the theory to its application. Stocks insulate subsystems from one another, and their exhaustion exposes dependencies. For the Persian Gulf states, restrictions on maritime supplies from spring 2026 are inadequately compensated by air and road transport alternatives. This depletes inventories unevenly across products and impairs the functioning of the logistics system itself. 
    Throughput optimisation builds fragility everywhere; the Gulf is where it is becoming visible, with the Strait of Hormuz restriction as the trigger. The shock is exogenous; the fragility it meets is not.
Formulated as a Kelly portfolio allocation under a weakest-link payoff, the optimal strategy concentrates resources on the lagging rather than the leading inputs.

\end{abstract}

\section{Introduction}\label{sec:intro}
\noindent
The term \emph{Kessler syndrome}~\citep{Kessler1978} describes a scenario where orbital object density crosses a critical threshold, allowing a random fragmentation event to trigger a runaway cascade and resulting in an altered final state.
The result is an impaired orbit. Atmospheric drag clears lower orbits, but higher orbits have no such clearing mechanism, and unless there is intervention they stay inaccessible\footnote{Permanent inaccessibility stems from an operational asymmetry: removing a satellite post-collision costs orders of magnitude more than deploying it, so a private asset becomes a systemic liability that no viable insurance premium can price. Michael Mainelli’s junk-bond proposal---higher upfront capital requirements on high-risk designs---uses financial screening to block reckless operators before the feedback loops trigger, though it ultimately relies on a sovereign backstop to absorb the unchargeable tail-risk.}.

\noindent
 This paper introduces a similar feedback structure for supply chains, where the absence of a clearing mechanism systematically  hampers operations. 
 Shortages become critical when they impair the machinery needed to replenish supplies, not when shelves are merely empty.
  A missing tyre removes a truck; a missing filter disables a generator; unavailable refrigerant impairs cold storage. These localised losses make the replacement parts required to repair the trucks, generators, and refrigeration systems themselves harder to deliver. While the original disturbance may come from outside, the amplification occurs inside, and  the consequence can be a qualitative jump in cost. Once a pipe network has ruptured in sufficient places, maintenance  turns into rebuilding cost, which is orders of magnitude higher.
   In the Gulf application there is a sharper variant of the same loop, since the mechanics and technocrats who install the parts and supervise the infrastructure are themselves an imported input with varying exit options.

\noindent
 The  changing nature of the feedback loop at the critical juncture is what unites both cases. In orbit, collisions create fragments that cause further collisions, accumulating debris. In supply networks, localised failures remove the exact capacities required to contain further failures, causing functional capability to disappear. 
 As in high orbit, where salvage must come from outside, a supply network trapped in this loop is blocked in practice at first, and in principle once the process has run long enough. The two sides of the blockage are the cash flow that covers routine operation but not reconstruction, and the physical means that the loop has itself been consuming. External commitment, often sovereign, can supply both, but only while the damage has not yet become irreversible. Beyond that point, neither money nor throughput restores the network. The term \emph{generalised Kessler syndrome} names this structure. In the Gulf states it would take the form of an inability to maintain the service bundle that advanced economic hubs require.




\noindent
The cause is structural: just-in-time inventory reduction~\citep{Womack1990}, single-sourcing, offshoring, and the elimination of redundant repair capacity convert buffers into working capital and concentrate supply.  
Costs are documented: disruption severity scales with network density and node centrality~\citep{Craighead2007}.
Optimisation of throughput under stable conditions, without reliance on diversification, removes slack (inventory, redundancy, multi-sourcing, in-house repairs), which  is a buffer against correlated shocks. 
A refutation would be a system\footnote{
Antifragility is a complementary idea, since  antifragile systems gain from volatility supported by optionality and redundancy. Throughput-optimisation under stable conditions strips this out.} in which coupling density, observation coverage, and repair capacity remain stable or improve as throughput expands.

\noindent
None of this is specific to the  Gulf,  
but this is where the mechanism is most glaring.
The chokepoint is mappable, the economies are dramatically dependent on external inputs, and results are visible. 
What the data show is the interval, not the outcome. Beneath the legible surface there may be an undercurrent: an economy still running beyond the edge, held up by buffers rather than flows. If so, it would be a Wile E. Coyote moment.
Failure can be latent. Recognition, when it comes, 
is instantaneous, as economic actors collectively realise\footnote{There is a gap between psychological recognition and the ability and the time needed for restocking or evacuation.} that the ground has disappeared.


\noindent
The theory comes next (\S\ref{sec:model}). Limited resource allocation  is a portfolio problem with a weakest-link payoff (\S\ref{sec:portfolio}). The application follows (\S\ref{sec:supply} and Appendix~\ref{sec:azzour}). 

\section{From Cascade Mechanism to Kessler Syndrome}\label{sec:model}
\noindent
A failure cascade exists when mishaps increase the number, severity or duration of incidents through internal dynamics. A  standard exogenous interruption exhausting an inventory at constant rate is depletion, not yet a Kessler syndrome. It becomes one when depletion changes the system so as to accelerate further depletion\footnote{Acceleration here means self-amplification---each failure raises the rate of further failures---not unbounded speeding-up. The cascade is bounded by the stock it consumes; what makes it Kessler is the altered final state, not the rate at which it is reached.} and eventually disable recovery.
What makes a supply crisis dangerous is weakest-link complementarity, not additivity. The output of a process depends on the simultaneous presence of every indispensable input and is therefore set by the least available of them; the absence of any one drops output to zero regardless of the abundance of the others.  Take a modern assembly line, where a missing microchip in a factory reduces synchronised labour and just-in-time materials from high throughput to pure overhead.

\subsection{Two Channels and one Amplifier}\label{sec:channels}

\noindent
    The opening paragraph named  the two steps of the Kessler transition common to satellites and supply systems---acceleration and loss of recovery.
Keeping them separate clarifies what is occurring in the Gulf. Two questions  transfer unchanged from orbit to the economy. 
Pre-threshold: which levers  define the tipping point—propagation/coupling, buffers, repair capacity, density? Post-threshold: is the resulting state hysteresis, or a qualitative change?


\noindent
\textbf{The additive channel.} A collision creates fragments; fragments strike
other satellites; those strikes create more fragments. Failure generates
failure, and the failure rate is self-amplifying until the process exhausts itself into a qualitatively  different state. The supply-chain analogue is
a shortage that causes a production failure, which causes a shortage elsewhere,
which causes further  production failure. The T\={o}hoku earthquake is the
canonical case~\citep{CarvalhoNireiSaito2014,Ivanov2019}.

\noindent
\textbf{The subtractive channel.} 
In orbit the subtractive loss comes from the disappearance of the satellites that provided collision avoidance, tracking and debris removal: the capacity that contains the hazard, destroyed by the hazard.  
Failure destroys the capacity to contain and repair failure, and the repair rate is degraded.
The supply-chain analogue is the mechanic who leaves, and the truck that fails because the part needed to fix it cannot be delivered\footnote{The additive and subtractive channels are analytically distinct but empirically entangled: the same event can do both. 
 As described in the main text, the cascade is dangerous because that capacity exists and can be removed. The distinction is a diagnostic for locating which channel is doing the work.
 Deterioration of infrastructure (weakening the dense web of connections and oversight) is subtractive; increasing chain length lengthens the path along which a failure propagates, which is additive.}.
 The original Kessler syndrome supplies the additive channel, while the generalised
supply-chain mechanism adds the loss of repair capacity as a second, distinct
channel. In orbit there is normally no internal repair capacity to destroy (repairs are organised from the ground post-event), so
the orbital case is mainly additive and supplies the analogy; the subtractive
channel is what the supply-chain case adds.

\noindent
\textbf{The structural amplifier.} More satellites in the same orbital volume
mean more collision paths; more processes in the same supply network, and
longer chains, mean more paths along which failures travel. The same density that makes the economy productive also makes it fragile.


\noindent
\textbf{Natural decay.} Atmospheric drag clears debris before it can cascade. In supply chains, redundancy, buffers, slack, and alternative suppliers absorb or delay shortages before they cascade. Throughput optimisation removes the natural decay; it is equivalent to moving to a higher orbit, where there is little drag. The decay asymmetry runs one way, and it decides what the post-threshold state looks like. This is also the answer to the objection that supply bottlenecks always self-heal.   Unmaintained infrastructure degrades by default, and functioning requires  maintenance and eventually substantial capital investment. Once the repair chain breaks, nothing remains functional for long, and the system relaxes to a lower complexity state rather than naturally returning to the pre-breakdown state\footnote{
Shelley's traveller finds Ozymandias's Works gone and only the trunkless legs still standing, as an ancient ruin survives as an artefact. Speer's ruin-value theory concedes  the point for  modern high maintenance architecture. Modern artefacts often leave  only indistinct rubble and a demolition order. 
}.
Risk-reduction strategies follow naturally. The initial interruption is exogenous and often hard to control; the other levers are controllable: redundancy and slack, shorter chains and less dependency, more diversification and alternative routes---more `space volume'. Each is the opposite of conventional throughput optimisation, none of them is free, and finding the minimal-cost solution for a required level of resilience is a permanent challenge.

\noindent
\textbf{Production.} Each service $j$ requires inputs indexed by $i \in E_j$. Output is
\begin{equation}\label{eq:oring}
y_j = \bar{y}_j \min_{i \in E_j} q_i, \qquad q_i \in [0,1],
\end{equation}
where $\bar{y}_j$ is installed capacity and $q_i$ is the availability factor of input $i$ ($1 = $ full supply, $0 = $ total stockout), defined in terms of stock in \eqref{eq:stock} below.  This is the Leontief fixed-proportions production function~\citep{Leontief1941} written in availability form: the limiting case of input complementarity~\citep{MilgromRoberts1995}, and the standard fixed-proportion structure of supply-chain and maintenance models. The output of a complex service is set by its least available indispensable input---a minimum, not a sum or an average, and not a product.\footnote{Multiplicative aggregation of input availabilities is deliberately not used. A product form prices independent Bernoulli error events---a task is ruined with probability $1-q_i$, independently across inputs---so \emph{expected} output factorises into a product of success probabilities. Physical stock availability is a state, not a trial: supplies are not like errors. The two forms agree on the catastrophe set (any $q_i = 0$ forces $y_j = 0$ in both) and coincide for binary availability states, but for partial availabilities the product understates feasible output, since $\min_{i \in E_j} q_i \ge \prod_{i \in E_j} q_i$ on $[0,1]^{|E_j|}$.} Production of a product can drop to zero when one essential part is missing.  The constraint is the lower tail of the availability of the essential inputs, not the mean.

\subsection{The Machinery of Failure}

\noindent
\textbf{Inventory dynamics.} The stock of input $i$ obeys
\begin{equation}\label{eq:stock}
\dot{s}_i = m_i + \rho_i + c_i - d_i - \omega_i,
\end{equation}
where each term is a flow in quantity per unit time: $m_i$ is effective delivery, $\rho_i$ local production, $c_i$ cannibalisation recovery (a transfer, not a new source), $d_i$ consumption, and $\omega_i$ spoilage. 
The availability factor is \(q_i = s_i / s_i^{\text{crit}}\) once \(s_i\) falls below the operating threshold \(s_i^{\text{crit}}\), and \(q_i = 1\) above it.
Nominal delivery is the delivery that would arrive if the transport service were fully available. Effective delivery is nominal delivery scaled by transport availability, a factor in \([0,1]\) equal to the weakest-link minimum over the transport service's indispensable inputs. Transport availability depends on the same trucks, power, maintenance and labour that the interruption degrades. Those inputs are treated as exogenous here.


\paragraph{Recovery is a Joint Condition.}
Repairing a failed service needs the part, a qualified technician, an
organisation able to hire him, a payment that clears, transport for both, and a
safe warehouse and road. Repair is itself production, so
\begin{equation}
r_j=\bar r_j\min\{q^{\rm stock}_j,\,q^{\rm labour}_j,\,q^{\rm env}_j\},\label{eq:repair}
\end{equation}
with $q^{\rm labour}_j$ covering both the technician and the organisation able to hire him, and  $q^{\rm env}_j$     itself a minimum over payment, transport and a safe warehouse and road.
Labour and stock are terms of one minimum, so interventions aimed at one term
are insufficient. 

\paragraph{Irreversibility.}
What is irreversible is not a part but the repair chain\footnote{An air-conditioning filter can sit in a warehouse while mould spreads
through the building it was meant to protect; the irreversibility came
from the joint failure, not from the filter's nature. The
same filter, with a technician, a contract, a payment, and a safe road, is a
reversible repair. What is condemned is the building, not the filter---and the
building is condemned because the repair chain failed.} failing for longer than
the damage it fails to contain. Let $T_{\rm abs}$ be the time for damage to
cross a condemnation threshold with no maintenance and $T_{\rm restore}$ the
time to restore the joint condition, both evaluated at the moment the joint condition fails.  The state is reversible iff
\begin{equation}
T_{\rm restore}<T_{\rm abs}.
\end{equation}
$T_{\rm restore}$ is days for a part, months for a technician or a payment
chain, and possibly years when the asset is condemned and capital must be raised; a
burst pipe or a biofilmed chilled-water loop is replaced, not repaired.

\noindent

\noindent
The weakest-link form does the work that a spectral-radius threshold would do in an additive model: it makes the system vulnerable to the exhaustion of a single indispensable input, not to a weighted average of failures. The cascade boundary is
a union of coordinate hyperplanes $\{s_i = 0\}$ (stockout hyperplanes of indispensable inputs); the system crosses it when any one coordinate hits zero. 
Substitution, rationing, and cannibalisation can postpone a crossing, but they cannot make an indispensable input dispensable. Only a restored flow lifts the minimum; without one\footnote{Substitution can significantly delay this outcome, as in the case of replacing a proprietary MED antiscalant with a generic less-efficient equivalent.
}, the crossing is a matter of time.

\noindent
\textbf{The input that is not a stock.} Parts have a reorder deadline; people do not.
Repair labour enters \eqref{eq:oring} like any other input---a plant with delivered parts but no (qualified) technicians has $q_{\text{tech}}$ of zero and output to match---but it obeys none of the inventory logic of \eqref{eq:stock}. It depletes not by consumption but by exit and by refusal to fill a role. If $\nu_k$ is the qualified coverage of repair function $k$, its dynamics are the pure-drain analogue of \eqref{eq:stock}, $\dot{\nu}_k = -x_k \nu_k$, with an exit hazard $x_k$ that rises with the current and the expected failure burden\footnote{ The exit hazard has an institutional component. European water companies, for example, rely heavily on specialised technical contractors in Gulf operations. These experts are engaged through commercial contracts rather than as employees. This substitution does not eliminate corporate duty of care; it changes its form—from a direct statutory employer duty to a contractual allocation of risk, priced and insured. The work proceeds; the nature of the liability shifts.}.
 A physical stock is backward-looking and declines smoothly; coverage is forward-looking and can jump: each departure raises the load on those who remain, and with it the hazard of the next departure---a run dynamic that depletion by consumption cannot reproduce.
This is the additive channel operating on the labour pool: failure generates
failure, applied to repair capacity itself. 
The classification of the mechanic as subtractive is therefore incomplete. While the loss is subtractive, the run that produces it is additive, and the run is what makes the labour channel non-linear in a way that inventory depletion is not. A stock declines smoothly; a pool can tip.

\noindent
Cannibalisation applies with the same debit logic as for parts---poaching a technician from a working facility strips the donor's coverage---with one asymmetry: the tacit, plant-specific knowledge a departing technician carries appears in no inventory, no manifest, and no customs category. 
The weakest link of \eqref{eq:oring} is usually read as a part; in the Gulf application the input at the minimum may name a person\footnote{This is related to Hirschman's work~\citep{Hirschman1970} on how employees express dissatisfaction.  For the expatriate technician, voice is largely unavailable and loyalty is contractual, so exit is itself the operative response.}.

\noindent
The framework distinguishes dynamic instability (failures propagate faster than repair contains them) from physical infeasibility (no feasible combination of inputs sustains the minimum bundle). The weakest-link structure makes the second sharper: the minimum bundle fails when any one of its indispensable inputs is exhausted.
The operative asymmetry is between observed flow and unobserved reserves. Water production and food prices attract attention; tyre compatibility, refrigerant inventories, filter condition, repair backlogs, and technician coverage may not. Stable output can coexist with shrinking $q_i$ in the lower tail, and a monitoring regime built on flows alone is structurally blind to half the trap. 

\noindent
The following conditional bound sharpens the distinction.
\begin{proposition}[Persistent essential deficit]\label{prop:deficit}
Fix a minimum service bundle. Let $V(t) \ge 0$ be a resource-equivalent reserve (accessible stocks plus salvageable components) that substitution and cannibalisation cannot increase. If every admissible policy maintaining the bundle satisfies $\dot{V} \le -\epsilon$ for some $\epsilon > 0$, the bundle cannot be maintained beyond $T=V(0) / \epsilon$.
\end{proposition}

\begin{proof}
Integrating $\dot{V} \le -\epsilon$ along any policy that maintains the bundle
gives $V(t) \le V(0) - \epsilon t$ for as long as the bundle is maintained. If
it is maintained throughout $[0,T]$, then $V(T) \ge 0$, so $T \le V(0)/\epsilon$.
\end{proof}

\noindent
The proof  follows directly from the assumptions. The work is in establishing the
premise: showing that every admissible policy draws $V$ down at a positive
rate requires commodity, substitution, bypass, and demand information. The
bound does not predict a date; it states what follows \emph{if} the premise
holds.

\noindent
The mechanism's defining feature is that it leaves no simple trace in flow data.
Flow data are consistent with its presence and with its absence; the
diagnostic series that could distinguish them (repair completion rates,
technician coverage, down-unit accumulation) are not public. The one clean test is retrospective: after the flow is restored, does the infrastructure return without new construction? If it does, the subtractive channel did not bind. If it does not, it did. Until then, the claim is a mechanism, not a diagnosis.

\noindent
The cascade-failure literature is vast. Interdependent networks exhibit first-order transitions when failure in one network loads another~\citep{Buldyrev2010}; load-bearing cascades on infrastructure graphs produce similar thresholds under redistribution rules~\citep{MotterLai2002,Kinney2005}; self-exciting point processes give a stochastic foundation with the branching ratio as a reproduction threshold~\citep{Hawkes1971,Brummitt2012}; firm-level propagation under input constraints has been simulated by~\citep{InoueTodo} and in dynamic input--output form by~\citep{Pichler}; the role of intermediate inputs and weak links in production is developed by~\citep{JonesC2011}, input specificity and shock propagation along supply chains by~\citep{BarrotSauvagnat2016}, and aggregate fluctuations from independent sectoral inventory dynamics by~\citep{Bak1993}.
The input-centric framing has deeper precedent in materials science: Weibull's weakest-link theory~\citep{Weibull1951} takes the flaw as the unit of analysis and yields the same weakest-link survival structure as a series system; Daniels's fiber bundle~\citep{Daniels1945} adds load redistribution as the cascade mechanism, the closest physical analog to repair rerouting. The supply-chain application is the translation of this tradition from flaw-to-input and from load-to-demand, not a new failure theory.

\noindent
Standard  models are node-centric---firms, sectors, or infrastructure assets as vertices, supplier--customer links as edges. The weakest-link framing here is input-centric: the unit of analysis is an indispensable input requirement of a process, not a firm or sector. It is complementary to the network approaches. It locates cascade risk in the weakest-link structure of the production function rather than in network topology. None of fixed-proportion complementarity, inventory dynamics, or cascade thresholds is new in isolation; the proposed contribution is the Kessler syndrome integration of an input-level weakest-link production function with inventory-mediated feedback, applied to a corridor disruption in which cargo composition matters as much as aggregate tonnage.

\paragraph{Many inputs, many clocks.}
A large process depends on many indispensable inputs, each with its own
run-out time. For independent inputs hazards add, so the bundle fails near the
lower tail of the marginals. 
Cargo moving through the Strait of Hormuz can be sorted by type, as examined in \citep{Meister2026}, with some categories passing through the chokepoint more easily as a group than others.

\subsection{Kelly Optimisation with Leontief Payoff}\label{sec:portfolio}
\noindent
Corporate financial optimisation and the structure of physical supply networks are symbiotic.
Working capital optimisation as well as Just-In-Time (JIT) manufacturing and zero-inventory management can be read as  operational expressions of the Kelly criterion~\citep{Kelly1956}. 
By purging idle cash drag and liquidating `wasteful' buffers, localised firms continuously maximise their capital growth rates. This relentless drive for efficiency is the primary engine of modern economic density and urban complexity.   The result is a structural trade-off rather than a failure of competence or systemic oversight, and it is governed by scale and diversification:
\[
\text{Macro Optimality} \;\Longrightarrow\; \text{Micro Fragility} 
\]
The macro-to-micro direction is a claim about the background model, not
a corollary of the water-filling result below. The reverse direction is less
natural: fragility does not imply optimality, since a budget can be wasted on
non-inputs, and a system can be fragile for reasons unrelated to any
optimisation at all. What that result supplies is the normative benchmark. It
shows what an optimal allocator would do with a given budget in the
deterministic setting, and the water-filling model is, in zeroth
approximation, a description of the state: its budget fixes the level it can
maintain, and whether that level suffices for the desired service bundle is
not self-evident.

\noindent
In classical portfolio theory, an aggregate allocator facing $N$
uncorrelated opportunities can eliminate idiosyncratic volatility by taking
infinitesimal position sizes $f/N$ as $N \to \infty$
(Arrow and Lind~\cite{arr}). Two features of the present setting weaken the
analogy. Output is governed
by non-additive Leontief minima ($\min_i q_i$) rather than linear portfolio
sums. Substrate dependencies enforce non-zero cross-node correlation.

\noindent
While the standard Kelly macro allocator can diversify across nodes, the nodes cannot diversify across their own inputs. The asymmetry is the point. The consequence is localised node collapse that does not necessarily show up as a failure at the larger scale. Throughput optimisation increases growth, but in the process it shifts additional risk onto the local nodes. Whether this macro- vs micro-level trade-off is viable remains an open empirical and social question.

\noindent
 \paragraph{Supply-chain Kelly.} The optimisation that follows is elementary---water-filling with a given budget $B$ as studied in Cover and Thomas~\citep{CoverThomas2001}---but its economic content is not. It formalises the inversion of the Kelly weighting, where capital that would flow to the strongest assets flows instead to the weakest indispensable inputs.  
Standard Kelly maximises $\mathbb{E}[\log(1+f\cdot R)]$ over an allocation $f$
with $\sum_i f_i = 1$. The optimum rewards an asset that pays off
when the rest of the portfolio does poorly, as well as assets with strongly favourable
returns.

\noindent
Replace the conventional payoff with the Leontief one, and the same log objective
inverts the weighting---capital now flows to the inputs below the water line,
and the inputs above it receive nothing. The criterion is Kelly's; the object
it rewards is inverted. Let the indispensable inputs sit at levels
$0 \le q_1 < \cdots < q_N$, let $f_i \ge 0$ be the resource allocated to
input $i$ with budget $\sum_i f_i \le B$, and let the service level be
$Z = \min_i (q_i + f_i)$.

\noindent
\begin{proposition}[Water-filling]\label{prop:waterfill}
Assume the $q_i$ are distinct. The allocation maximising $\log Z$ subject to
$\sum_i f_i \le B$ is water-filling. Let $S_k = \sum_{i \le k} q_i$ and let
$K$ be the unique index with
\begin{equation}\label{eq:K}
(K-1)\,q_K - S_{K-1} \;\le\; B \;<\; K\,q_{K+1} - S_K ,
\end{equation}
taking $q_{N+1} = +\infty$. Then
\begin{equation}\label{eq:zstar}
Z^{*} = \frac{S_K + B}{K},
\qquad
f_i^{*} = \max\{0,\, Z^{*} - q_i\}.
\end{equation}
\end{proposition}

\begin{proof}
For any achievable level $Z$, feasibility requires $f_i \ge Z - q_i$ for every
$i$, hence
\[
\sum_i f_i \;\ge\; C(Z) := \sum_i \max\{0,\, Z - q_i\}.
\]
$C$ is convex, piecewise linear, and nondecreasing, with
$C(Z) = k Z - S_k$ on $[q_k, q_{k+1}]$. The largest feasible $Z$ is therefore
the one solving $C(Z) = B$, which is \eqref{eq:zstar};  the index $K$ is fixed
by $Z^{*} \in [q_K, q_{K+1}]$, which is \eqref{eq:K}. The allocation
$f_i^{*} = \max\{0, Z^{*}-q_i\}$ uses the budget exactly and attains
$\min_i(q_i + f_i^{*}) = Z^{*}$. 
\end{proof}

\noindent
\begin{remark}[Availability cap]\label{rem:cap}
The solution assumes the budget can raise any input without bound, whereas
\S\ref{sec:model} has $q_i \in [0,1]$. If availability is capped at
$q_i + f_i \le 1$, the formula applies whenever $Z^{*} \le 1$; if the budget
is large enough that \eqref{eq:zstar} would exceed 1, then $Z^{*} = 1$, all
inputs are saturated, and the residual budget $B - \sum_i (1-q_i)^{+}$ is
idle. 
Proposition~\ref{prop:discrete} uses this
uncapped formulation; the numerical example below lies in a regime where
the cap is nonbinding.
\end{remark}

\noindent
Three  properties distinguish \eqref{eq:zstar} from standard Kelly.

\noindent
(i) \emph{Concentration}: in the deterministic case, only inputs below $Z^{*}$
receive resources; the rest receive nothing\footnote{Under uncertainty, the strict concentration result weakens:
any indispensable input with a positive probability of zero
availability must receive positive cover, since otherwise
$\mathbb{E}[\log Z]=-\infty$. If the limiting input is unique
almost surely at the optimum---that is, ties occur with
probability zero---the marginal value of cover for input $i$ is
then $\mathbb{E}[Z^{-1}\mathbf{1}_{\{i\text{ is limiting}\}}]$.
At an optimum, this marginal value is equal across funded inputs.
A general treatment of the stochastic problem via the KKT
conditions is beyond the scope of this paper. The two-input
binary case of Proposition~\ref{prop:discrete} is a solvable
instance.
}.

\noindent
(ii) \emph{Equalisation}: the active inputs are levelled at $Z^{*}$, not
weighted by their returns. The mechanism is to lift the weakest to a common
floor, not to reinforce the strongest.

\noindent
(iii) \emph{Added inputs lower the level}: an additional indispensable input
below the water line weakly lowers $Z^{*}$, since the budget must now also
lift the newcomer. 

\noindent
    This is the formal content of the observation that nodes cannot diversify across their own inputs: the growth-optimal response to a weakest-link payoff is to fund the bottom of the distribution, and adding indispensable inputs lowers the level that funding can reach.

\noindent
\paragraph{The state as allocator.} The supply-chain Kelly problem of
Proposition~\ref{prop:waterfill} is the state's challenge as much as any
firm's: given a budget $B$, it maximises the service level $Z^{*}$
achievable with the inputs at hand. The cost of a target bundle
$\bar Z$ is $C(\bar Z) = \sum_i \max\{0, \bar Z - q_i\}$; if $B < C(\bar Z)$,
the bundle fails. The budget $B$ here is physical throughput, not abstract
fiscal capacity. A state can have a liquid sovereign wealth fund, but not
containers arriving in port. 
Its vulnerability is the same as any throughput
optimiser's: $B < C(\bar Z)$, borne at the scale of a sovereign
economy. 

\noindent
\begin{proposition}[Discrete weakest-link allocation]\label{prop:discrete}
Two indispensable inputs, indexed $i \in \{1,2\}$, have independent binary
availabilities: input $i$ sits at level $q_i > 0$ with probability
$p_i \in (0,1)$ and at level $0$ otherwise. The split $f_1 = t$,
$f_2 = B - t$, with budget $B > 0$, is chosen before the states are observed, and growth is
$\Phi(t) = \mathbb{E}\ln z(t)$, with $z(t)$ the realised minimum of the two
levels. Then,
\begin{enumerate}
\item[\emph{(i)}] \emph{Piecewise quadratics.} $\Phi$ is concave and
piecewise smooth; its kinks, when they occur in $(0,B)$, are at
 $t \in \{\tfrac{B-q_1}{2},\, \tfrac{B}{2},\, \tfrac{B+q_2-q_1}{2},\,
\tfrac{B+q_2}{2}\}$, where the four levels
 $q_1+t$, $t$, $q_2+B-t$, $B-t$ cross pairwise. 
 On each piece at most three distinct levels bind, their probabilities sum to one, and stationarity reduces to a polynomial of degree at most two.
 When three levels bind, the quadratic coefficient is \(\pm1\).
\item[\emph{(ii)}] \emph{Symmetric case.} If $q_1 = q_2 = q$ and
 $p_1 = p_2 = p$, then $t^{*} = B/2$ and
 $\Phi(t^{*}) = p^{2}\ln\!\bigl(q + \tfrac{B}{2}\bigr) +
(1 - p^{2})\ln\!\bigl(\tfrac{B}{2}\bigr)$.
\end{enumerate}
\end{proposition}

\begin{proof}
(i) $z$ is a minimum of affine functions of $t$, so $\Phi$ is concave. At
$t=0$ input 1's down-level is $0$, so $z=0$ with probability $1-p_1>0$ and
$\Phi(0)=-\infty$; likewise $\Phi(B)=-\infty$. Hence the maximum is interior,
$0<t^*<B$, and the binding pattern is constant between level crossings,
i.e.\ between the four displayed thresholds. A pattern in which all four levels bind is impossible:
If one input’s up-level exceeds the other input’s up-level, it cannot be the minimum in any state, since the other input’s realized availability never exceeds its own up-level. Hence at most one up-level and the two down-levels can bind. Hence
at most three levels bind; writing them $\alpha_j + \sigma_j t$,
 $\sigma_j = \pm 1$, with probabilities $\pi_j$ summing to one, stationarity
reads $\sum_j \pi_j \sigma_j / (\alpha_j + \sigma_j t) = 0$, and clearing the
three denominators leaves a quadratic with $t^2$-coefficient
 $(\pi_1+\pi_2+\pi_3)\,\sigma_1\sigma_2\sigma_3 = \sigma_1\sigma_2\sigma_3 = \pm 1$,
 the product of the slopes of the three binding levels.
(ii) Exchangeability gives $\Phi(t) = \Phi(B - t)$, so concavity puts the
optimum at $B/2$; the states then deliver $z = q + B/2$ when both inputs are
up (probability $p^2$) and $B/2$ otherwise.
\end{proof}

\noindent
\emph{Example (asymmetric reliability).} Let $q_1 = q_2 = 0.4$, $B = 0.2$,
 $p_1 = 0.9$, $p_2 = 0.5$. The middle thresholds coincide at $t = 0.1$. On
 $(0, 0.1)$ the binding levels are $q_1+t$, $t$, $B-t$, and stationarity is
\[
\frac{p_1 p_2}{q_1 + t} + \frac{1 - p_1}{t} = \frac{p_1(1-p_2)}{B - t},
\qquad\text{i.e.}\qquad t^2 + 0.11\,t - 0.008 = 0,
\]
with root $t^{*} = 0.05$. On $(0.1, 0.2)$ the binding levels are
$q_2+B-t$, $B-t$, $t$, and the first-order condition there reads
\[
\frac{p_1 p_2}{q_2 + B - t} + \frac{1-p_2}{B-t}
= \frac{(1-p_1)p_2}{t},
\]
whose left side exceeds the right side throughout the interval, so $\Phi$ is
decreasing there and the maximum on that piece is at the kink $t = 0.1$.
Hence the optimal split is $(0.05,\, 0.15)$. The equal split attains
$\Phi = -1.578$ against $-1.513$; water-filling on the mean availabilities
$(0.36, 0.20)$ gives $(0.02, 0.18)$ and $-1.553$; and both endpoints are
ruin. The budget flows three-to-one to the \emph{less} reliable input,
because its down-state is the modal catastrophe---it sets the minimum with probability 
$p_1(1-p_2) = 0.45$---and the stationarity condition weights states by probability times inverse service level; this product is largest in that modal state. 
For $N \ge 3$, the coverage conclusion persists while allocation becomes
a multidimensional concave optimisation problem; the two-input case above
provides an explicit illustration. Uncertainty thus amends property~(i) of
Proposition~\ref{prop:waterfill} in the expected direction: every link with
a down-state is funded at any positive budget, if possible, because the log's
infinite penalty for ruin is what enforces it. 

\noindent
The German army logistics crisis during the 1941 Russian winter~\citep{MGFA2015} supplies a different example of an `optimal' allocation mechanism that failed. Winter clothing arrived late or never amid a transport impasse. Ammunition and fuel received priority  while bulky clothing waited, and the soldiers' ability to use those supplies deteriorated. 
The missing consumable destroyed the soldiers' own capacity---a state variable, not a stock---to use the supplies that did eventually arrive.
Bulky, commonplace consumables are perhaps the modern analogy to winter clothing: individually low-salience, but disastrous when missing at scale.
\noindent
The German army fought for three more years, since inertia is real, and humans improvise. But the war machine was  severely impaired by the logistical failure.
\noindent

\section{The Mechanism is not specific to the Gulf}\label{sec:scope}

\noindent
The Gulf is in the news, and three features make the crisis unusually
legible: the chokepoint is a single, mappable geographic gate; the disruption
is fast enough that the sequence from buffered interruption to recovery
erosion unfolds within months; and the affected economies' 
 import dependence is concentrated enough that the
aggregate statistics move visibly. All are properties of the observation, not of the mechanism.

\noindent
 The same structure is present in supply systems unrelated to the Gulf.
Centralised sewage treatment is a Leontief process with a small indispensable
input set and little redundancy at the catchment scale: a single failure in a pump
station could threaten a system
serving a city if not caught by regular maintenance and  timely repairs.
Sealed
building envelopes, designed for thermal efficiency, couple habitability to
continuous mechanical ventilation and cooling; within a couple of days of losing
power and water, the envelope that saved energy becomes the reason the
interior is uninhabitable. 

\noindent
Of course, few such systems are on the edge of collapse, because the external flow is reliable or easily restored before the binding margin expires. The argument is that the \emph{class} of failure is general, that its early stages are invisible in flow data (the asymmetry of observed flow versus unobserved reserve, \S\ref{sec:model}), and that the Gulf compresses the timescale enough to see the sequence.


\section{The Gulf Supply System}\label{sec:supply}

\noindent
The Gulf Cooperation Council (GCC) states have varying sea access, and are not all solely reliant on the Persian Gulf. Saudi Arabia has Red Sea ports. Oman and UAE have direct access to the Indian Ocean. Kuwait, Bahrain, and Qatar differ in land connections and onward routing.

\begin{figure}[tbp]
    \centering
    \includegraphics[width=0.5\columnwidth,keepaspectratio]{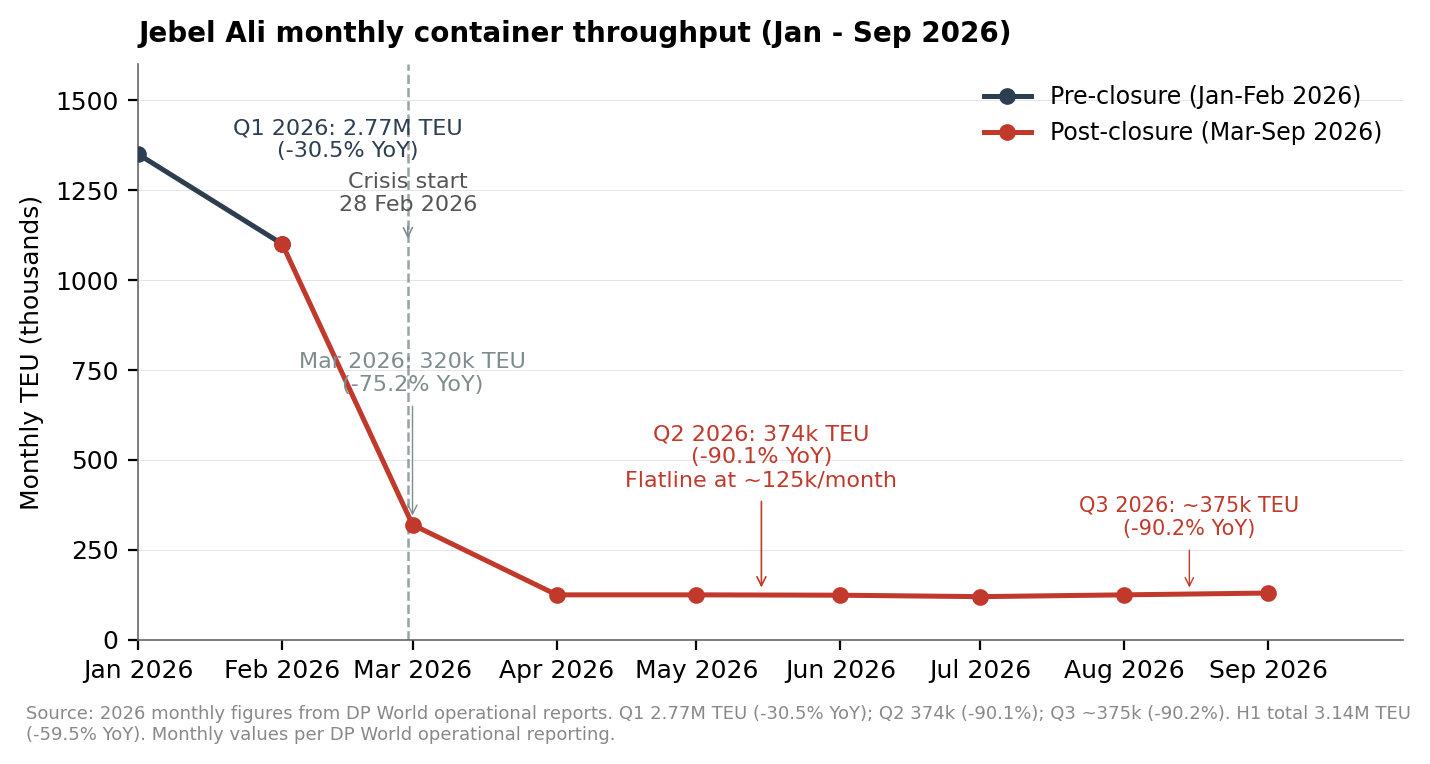}
    \caption{Jebel Ali monthly container throughput, January–September 2026. Pre-closure baseline $\sim1.1$--$1.35$ M TEU/month;
February 2026 disruption onset; contraction to $\sim 125$k TEU/month from April onward ($\sim$90\% YoY decline). Q1 2026: 2.77~M TEU ($-30.5\%$); Q2 2026 reported
total: 374k TEU ($-90.1\%$). Source: monthly reconstruction from the quarterly figures in DP World's 13~August 2026 release~\citep{DPWorld2026} (Q1--Q2 reported); months after June are indicative.
    }
    \label{fig:jebel_ali}
\end{figure}

\noindent
\paragraph{Jebel Ali.} Evidence of severe disruption at a major gateway exists. DP World's 13~August 2026 results~\citep{DPWorld2026} report Jebel Ali throughput of 374{,}000~TEU in Q2, down 90.1\% year-on-year (first half $-59.5\%$), alongside expanded inland connectivity. Maersk's 17~September update~\citep{Maersk2026} documents booking restrictions, landbridge options, and cargo-specific exceptions. 
Figure~\ref{fig:jebel_ali} shows a monthly reconstruction for January--September 2026, derived from the cited quarterly figures (Q1--Q2 reported; later months indicative): a pre-disruption January--February baseline of $\sim1.1$--$1.35$~million TEU/month, a structural break at the 28~February 2026 conflict onset, a 75\% year-on-year drop in March, and a flat $\sim$125{,}000~TEU/month (roughly 10\% of the pre-disruption level) from April through September, with no recovery in the quarterly reported totals. The reported quarterly figures support a large Q2 decline.
Appendix~\ref{sec:azzour} reports facility-level satellite signals at Az-Zour North One showing no decline through October 2026, consistent with the first two stages of the sequence.

\noindent
\textbf{The entry and exit channel for Labour.}  Labour is a state variable, subject in the GCC to the    Kafala system and its modern adaptations. The state variable is large in the Gulf with  a migrant population of thirty-five million~\citep{AlJazeera2026}, and they are the majority  in Qatar, the
UAE, and Kuwait. The channel opens before the worker arrives: the visa names his
profession and his sponsor, and oversight passes to a
private party at the border itself. 
The system is not uniform,
and the variation between GCC countries is substantial. 
Formal sponsorship rules differ across the GCC and have been loosened in several states, but the changes have often been more substantive on paper than in practice: employers still control residency status, including cancellation and absconding reports. For the model the legal detail matters less than two practical facts. An employer can end a technician’s local professional existence by cancelling or not renewing his residency, so repair capacity can be shed administratively, 
while workers with options can leave on a whim.  
For an elevator mechanic, who cannot easily resign and switch jobs, the balance of power is skewed towards the employer. His visa can be cancelled or simply not renewed, his file closes. 
He might still be in the city, part of the informal economy, but he is no longer an elevator mechanic hireable by an authorised firm   and traceable by a procurement system.
His disappearance silently removes skills, leaves infrastructure unmaintained, and damages economic viability. An elevator stays broken while its spare parts sit in a crate; offices become unreachable.
This is the dog that didn't bark. The system did not notice the loss because it took the mechanic's services for granted.

\noindent
\textbf{The transport-interruption feedback loop.} The
transport that carries supplies is itself maintained by them. Draining the
stock degrades transport, and degraded transport caps rebuilding. When external
access is at or below normal consumption, nothing rebuilds. When access exceeds
normal consumption, rebuilding requires a cover high enough that the degraded
transport can still carry the recovery. The feedback loop binds twice: the transport
that drained the stock also caps its rebuilding.

\noindent
Four overlapping stages of the transition can be distinguished, with the
subtractive channel dominating from the third: (1) \emph{buffered interruption}---overall imports fall, but stocks and alternative routes preserve service; (2) \emph{selective run-out}---some complementary inputs  are scarce,  while substitution and cannibalisation maintain visible output at the cost of reserves; (3) \emph{recovery erosion}---logistics and utility failures reduce effective deliveries and repair capacity; shared queues grow, and recovery is capped by the degraded
transport that drained the stock; (4) \emph{loss of the service bundle}---no
feasible allocation maintains the bundle associated with the state as an
advanced economic hub, and the system relaxes to a simpler
configuration.  The stages are a diagnostic tool, not a deterministic road map. Direct damage can make a system skip a stage, while rapid trade recovery, supported by emergency financing, can 
allow the system to backtrack. 

\paragraph{The Latent State.}
Eight months without manifest failure is what the framework expects of its
first two stages, buffered interruption and selective run-out. It is compatible
with deep stocks, with rerouting  of  essential parts, with fewer
external inputs (reactive maintenance, substitution, cannibalisation), and with
reserves being consumed unobserved: infrastructure condition, repair backlogs,
technician coverage. Aggregate port
volumes cannot give answers; granular customs and procurement records, maintenance
backlogs and administrative labour series can.

\paragraph{Trade Flow through the Strait of Hormuz.}\label{sec:hormuz}
The companion paper~\citep{Meister2026}\footnote{The earlier paper lacks a clear statement about the value-density difference between commodities and the vessels that carry them. A multi-million dollar toll is more easily absorbed by a high-value VLCC payload  than by bulk aggregates travelling on smaller ships worth a fraction of a tanker's cargo value. Covering transaction costs from a smaller financial base is proportionally harder. 
Driven by earlier sanctions, crude oil  relies on an established `shadow fleet'; LNG must be strictly traceable to its source, making clandestine blending or ship-to-ship (STS) bypasses commercially impractical, while sulphur or limestone cannot be transferred at sea without extensive transshipment infrastructure.  } examined the Strait of Hormuz. The enforcer's attack intensity and toll schedule force the follower to make a choice on whether imports through a toll-requiring route can be sustained. Container lines face higher transaction costs than tankers.
Bulk imports and especially oil flows have held up better and have at times recovered to pre-crisis levels. The two papers are complementary.

\section{Conclusion}\label{sec:conclusion}
\noindent
The paper introduced the generalised Kessler syndrome and an application to supply chains of the Gulf States.  Growth-optimisation can  
reduce resilience, while inventories---the activation mechanism---intensify the system coupling during the crisis itself. The Gulf suggests a plausible sequence: maritime interruption, uneven depletion, failure of inconspicuous complements, loss of logistics and maintenance, reduced capacity to correct further shortages. Cannibalisation, fungibility, and prioritisation of emergency imports alter the sequence and delay failure. The paper has described a plausible mechanism (Proposition~\ref{prop:deficit}) and, in its portfolio form (Proposition~\ref{prop:waterfill}), the inversion of the Kelly weighting: growth-optimal allocation under a weakest-link payoff concentrates on the lagging inputs. 
What would confirm the Kessler syndrome framing for a given state, and not just a cascade, is a documented feedback from shortage to lost function.  The emergence of a district of uninhabitable buildings, or a  whole network of burst pipes, would be an irreversible event. 

\noindent
Sovereign state economies are exceedingly complicated and the mechanism  presented in Section~2 only captures the general phenomenon; when and how a particular state will have to radically simplify is left for future research.


\noindent
 From outside the region, the episode invites a comfortable, even smug, reading: a set of states with a particular weakness, exposed when a chokepoint closed.
 The theory resists this interpretation. 
 Throughput optimisation converts slack into coupling wherever it runs, and the fragility it builds is a global condition that happened to be triggered at Hormuz.

\noindent
A mundane example is the Danish water treatment sector, which transformed decentralised, high-redundancy municipal plants into tightly coupled regional hubs\footnote{To the municipal planners and environmental engineers, replacing hundreds of weakly monitored local sites with a handful of centralised, highly instrumented facilities seemed an operational triumph. Yet in optimising for ease of oversight and throughput efficiency, they  converted network slack into tight coupling, leaving large urban catchments dependent on the uninterrupted performance of a few critical nodes.
}; the  same pattern is replicated
across the world. 
 Any sufficiently concentrated interruption---an earthquake, a strike, a bankruptcy, a cyber event, a different chokepoint for a different flow---supplies the initial condition.

\noindent
Another striking example is the global stock of sealed, mechanically ventilated high-rise buildings, which has compressed the habitability horizon following power-grid or water failure to $24\text{--}72$ hours, and to only a few hours for all-glass envelopes under extreme heat\footnote{The New York City 1977 blackout ran $25$ hours in a July heatwave. The city came through because power returned and its older residential stock---operable windows, heavy masonry---was passively habitable, which is the comparison that matters here. Sealed glass is not: with the grid down, standby diesel is sized for life-safety loads, not cooling, so the interior becomes uninhabitable within a few hours.}.
Modern sealed high-rises convert energy and civil engineering efficiency into acute systemic vulnerability during blackouts\footnote{This  habitability threshold inventory is substantial. There are about 35-40 completed `supertalls' (300m+, roughly 80-100+ stories) in the GCC out of around 250 worldwide, and in addition tens of thousands of mid-to-high-rise sealed envelopes. The global stock of mechanically dependent high-rises exceeds $100,000$.
Supertalls highlight a structural asymmetry and exhibit a reversibility challenge: assembly is an  integration of standardized components into uniform geometry, whereas dismantling must reverse path-dependent variation. 
Civil engineering  prioritises assembly  and supertall deconstruction is  untested.
An economic asymmetry also exists, between financing a new tower and taking one down.
}.

\noindent

\noindent

\noindent
This fragility of cities or whole civilisations is not new.
Even ancient societies  relied on  long-distance trade for vital resources: salt, copper,  tin, and iron. What has changed is the temporal horizon of vulnerability. Earlier disruptions\footnote{The collapse of the Bronze Age trade routes under pressure of the mysterious Sea People would be such an example.} unfolded
over months or years; the modern sealed skyscraper compresses that margin down to a few days. This risk should not be viewed in isolation, since it is counterbalanced by benefits.  The modern sealed skyscraper is the latest   version of this Faustian bargain.

\noindent
While the Kessler syndrome is abstract, the Gulf is a real-time experiment of novel proportions in size and speed\footnote{One of the technical novelties of the current disruption lies less in for example the interception bottlenecks than in the radical collapse of Circular Error Probable (CEP). 
Lethal effect per unit area decays non-linearly with distance---the areal
density of fragments from a burst follows an inverse-square law
($I(r) \propto r^{-2}$)---and hit probability against a point target scales
roughly with the inverse square of the guidance error. Reducing terminal
guidance error by an order of magnitude therefore multiplies
breach probability against static assets by up to two orders of magnitude.
Consequently, the cost of passive capital hardening (such as reinforcing `every' aircraft hangar) grows uncontrollably, turning traditional static defense into a mirage, with no structural inversion of the cost curve in sight.}. 
A specific component does not characterise it, since spare parts and the elevator mechanic mentioned before are just links of a failure chain.
The countdown clock is what matters. Once started, it runs at a situation-dependent speed\footnote{Historical collapses ran on slower clocks. In the late Bronze Age collapse ships  carried both the threat and the warning.}, and has an outcome that is irreversible unless the flow is restored before the damage crosses the threshold.
In the Gulf at least three clocks run at once: information is instant, failure takes days, while a corrective response takes months.  The warning outruns the threat, but nothing else does.
\noindent
The mechanic who went missing is the final unheard messenger. 
His departure appeared in no flow statistic and cannot be reversed by any shipment. Abruptly, the system stops.


\paragraph{Data \& Tool Attribution.}
The paper uses open-access satellite imagery from Copernicus Sentinel-2  and the USGS/NASA Landsat programme, processed with the Google Earth Engine Python API; data aggregation, analysis, and visualisation with the open-source Python libraries GeoPandas, Pandas, Matplotlib, and Geemap.

\appendix
\section{The Az-Zour North One Desalination Plant  }\label{sec:azzour}
\noindent
Desalination plants reliant on a range of technologies  are essential for all Gulf states. Here data is shown for  Az-Zour North One, which seems to have fared well during the crisis.
Az-Zour North One links electricity and water production in a single  facility.
The water process is multi-effect distillation (MED); it differs from the reverse-osmosis process at Az-Zour South~\citep{ShamalAzZour,Veolia}.

\begin{figure}[tbp]
    \centering
    \begin{minipage}[t]{0.49\columnwidth}
    \captionsetup{type=figure}
        \centering
        \includegraphics[width=\linewidth,keepaspectratio]{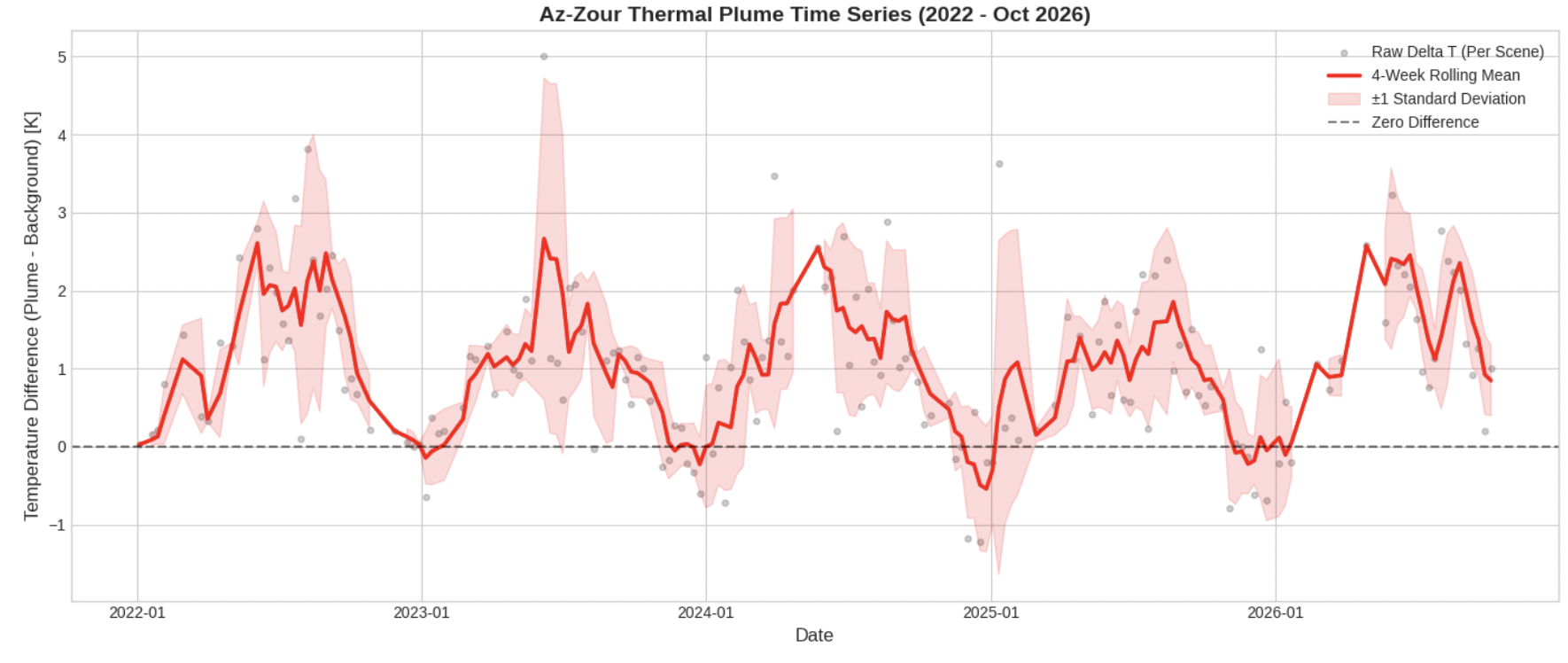}
        \caption{Az-Zour North One Thermal Plume Time Series (2022--Oct 2026). Grey points are labelled per-scene observations; the red line is labelled a four-week rolling mean and the band $\pm 1$ standard deviation. The vertical axis is a plume-minus-background index difference. Source data: LANDSAT/LC08/C02/T1\_L2~\citep{LandsatST}.}
        \label{fig:thermal}
    \end{minipage}\hfill
    \begin{minipage}[t]{0.49\columnwidth}
    \captionsetup{type=figure}
        \centering
        \includegraphics[width=\linewidth,keepaspectratio]{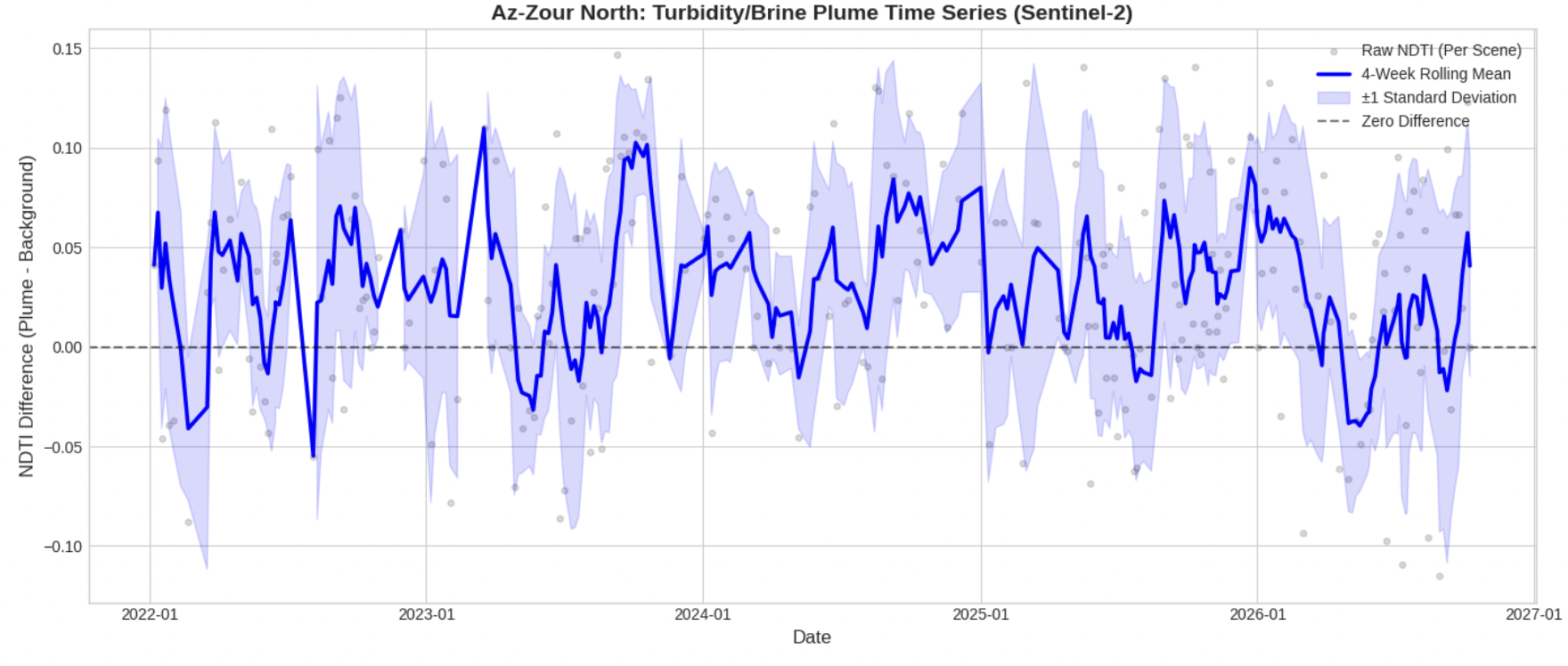}
        \caption{Az-Zour North One Turbidity Plume Time Series (2022--Oct 2026).
        Grey points are labelled per-scene observations; the blue line is labelled a four-week rolling mean and the band $\pm 1$ standard deviation. The vertical axis is a plume-minus-background index difference.
        Source: Copernicus Sentinel-2 NDTI time series~\citep{Sentinel2}. }
        \label{fig:optical}
    \end{minipage}
\end{figure}

\noindent
Figure~\ref{fig:thermal}\footnote{During part of the winter and spring periods,  cloud cover can reduce valid per-scene observations below the threshold required to compute a  standard deviation over the 4-week rolling window, resulting in a temporary lapse of the variance band.} plots the temperature difference between a designated plume area and background water from 2022 to October 2026 using USGS Landsat 8 data.
 A low thermal contrast can reflect changed discharge, ambient conditions, or mixing;  converting the thermal index to changes in plant output requires calibration.
 
\noindent
Figure~\ref{fig:optical} shows a Sentinel-2 NDTI difference between plume and background areas. The turbidity contrast is intermittent and at times negative. This supports the inclusion of an optical (turbidity) channel alongside temperature in a monitoring programme. 
The figure label `Turbidity/Brine Plume' should not be read as a properly calibrated salinity measurement: the index is an optical proxy whose relation to suspended material, discharge and local water conditions has not been established. It requires further analysis and validation.
A comparison with other plants of all types\footnote{MED, reverse osmosis (RO), and multi-stage flash (MSF) being the major types.} across the region would add clarity, but is not included here.

\begin{figure}[tbp]
    \centering
    \includegraphics[width=0.72\columnwidth,keepaspectratio]{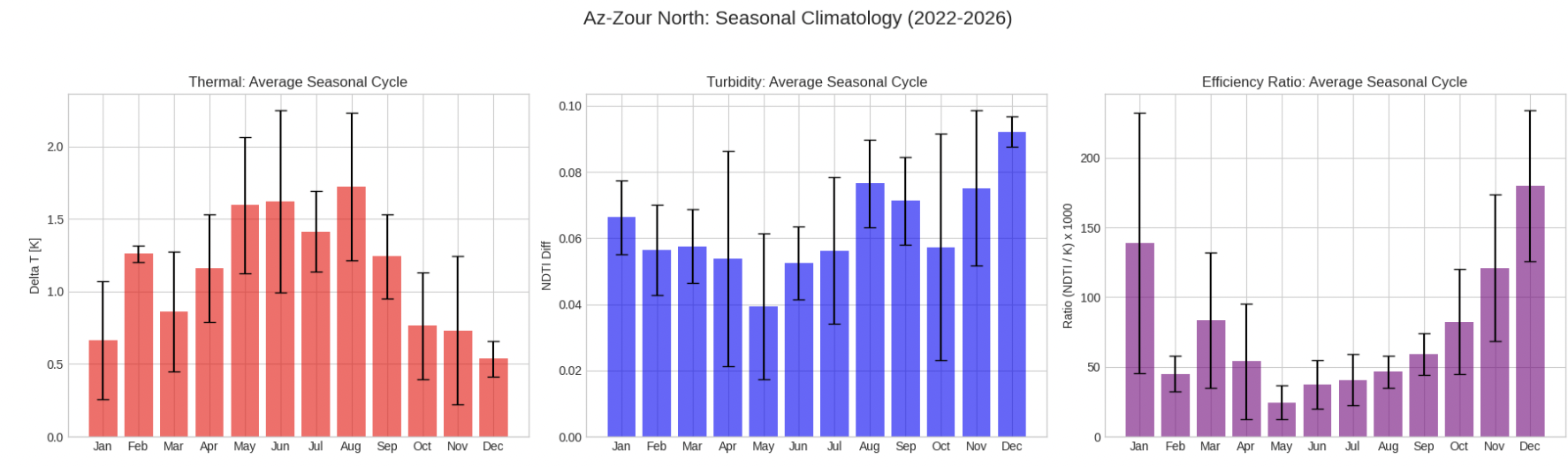}
    \caption{Az-Zour North One seasonal climatology (2022--Oct 2026). Monthly means and ranges of the thermal contrast (left), the turbidity contrast (centre), and their ratio (right), aggregated over the plotted period. Source: See Figures~\ref{fig:thermal} and~\ref{fig:optical}.}
    \label{fig:seasonal}
\end{figure}

\noindent
Figure~\ref{fig:seasonal} makes the seasonal structure visible. The thermal contrast peaks in summer and troughs in winter, the turbidity contrast varies less, and their ratio therefore peaks in winter, when the thermal denominator is smallest.
The first two plots  show 
   no obvious sustained discontinuity in the plotted signals.

\FloatBarrier
\begin{multicols}{2}
 
\footnotesize
\setlength{\bibsep}{0pt}

\end{multicols}

\end{document}